\documentclass[conference]{IEEEtran}
\IEEEoverridecommandlockouts
\usepackage{cite}
\usepackage{amsmath,amssymb,amsfonts}
\usepackage{graphicx}
\usepackage{textcomp}
\usepackage{xcolor}
\def\BibTeX{{\rm B\kern-.05em{\sc i\kern-.025em b}\kern-.08em
    T\kern-.1667em\lower.7ex\hbox{E}\kern-.125emX}}

\usepackage{algpseudocode}
\usepackage{algorithm}
\usepackage{relsize}
\usepackage{mathtools}
\usepackage{multirow}
\usepackage{comment}
\usepackage{listings}
\usepackage{enumitem}
\usepackage{subcaption}
\usepackage{url}
\usepackage[hidelinks]{hyperref}

\usepackage{pifont}

\newcommand\inlineCPP[1]{\lstinline[basicstyle=\small\ttfamily,language=c++]|#1|}
\newcommand{\CC}{C++}
\newcommand{\PP}{Pregel+}

\definecolor{mygreen}{HTML}{1B5E20}
\definecolor{myred}{HTML}{d62728}

\lstdefinelanguage{palgol}{
 morekeywords={for,in,do,until,fix,repeat,exists,forall,end,if,else,
  let,true,false,fst,snd,to_int,to_float,ref,val,inf,local,remote,V,
  input,output,field,extern,maximum,minimum,sum,and,or,random,
  Nbr,In,Out,Id},
 sensitive=true,
 morecomment=[l]{//},
}
\lstdefinestyle{highlight}{
 basicstyle=\scriptsize\ttfamily,
 language=c++,
 morekeywords={override},
 numbers=left,
 stepnumber=1,
 firstnumber=1,
 xleftmargin=2.5em,
 commentstyle=\color{mygreen},
 moredelim=**[is][\color{myred}]{@}{@}
}

\begin{document}
\title{\Large \bf Composing Optimization Techniques for Vertex-Centric \\ Graph Processing via Communication Channels
}

\author{
\IEEEauthorblockN{
Yongzhe Zhang\IEEEauthorrefmark{1}\IEEEauthorrefmark{2},
Zhenjiang Hu\IEEEauthorrefmark{1}\IEEEauthorrefmark{2}\IEEEauthorrefmark{3}}
\IEEEauthorblockA{\IEEEauthorrefmark{1}Department of Informatics, SOKENDAI, Japan}
\IEEEauthorblockA{\IEEEauthorrefmark{2}National Institute of Informatics (NII), Tokyo, Japan}
\IEEEauthorblockA{\IEEEauthorrefmark{3}Department of Information and Communication Engineering, University of Tokyo, Japan}
}

\maketitle

\begin{abstract}
Pregel's vertex-centric model allows us to implement many interesting graph algorithms, where optimization plays an important role in making it practically useful.
Although many optimizations have been developed for dealing with different performance issues,
it is hard to compose them together to optimize complex algorithms, where we have to deal with multiple performance issues at the same time.
In this paper, we propose a new approach to composing optimizations, by making use of the \emph{channel} interface, as a replacement of Pregel's message passing and aggregator mechanism, which can better structure the communication in Pregel algorithms.
We demonstrate that it is convenient to optimize a Pregel program by simply using a proper channel from the channel library or composing them to deal with multiple performance issues.
We intensively evaluate the approach through many nontrivial examples.
By adopting the channel interface, our system achieves an all-around performance gain for various graph algorithms.
In particular, the composition of different optimizations makes the S-V algorithm 2.20x faster than the current best implementation.
\end{abstract}


\section{Introduction}
\label{sec:introduction}

Nowadays, with the increasing demand of analyzing large-scale graph data in billion or even trillion scale (e.g., the social network and world wide web), lots of research~\cite{survey, survey2} has been devoted to distributed systems for efficiently processing large-scale graphs.
Google's Pregel system~\cite{pregel} is one of the most popular frameworks to handle such kind of massive graphs.
It is based on the BSP model~\cite{bsp} and adopts the vertex-centric paradigm with explicit messages to support scalable big graph processing.
Pregel's vertex-centric model has demonstrated its usefulness in implementing many interesting graph algorithms~\cite{pregel, QuWH12, connectivity, optimizing},
and imposed influence over the design of Pregel-like systems, such as Giraph~\cite{giraph}, GPS~\cite{gps}, Mizan~\cite{mizan}, \PP{} \cite{pregelplus}.

While Pregel provides a friendly interface for processing massive graphs,
current research shows that it is important to introduce optimizations for dealing with various performance issues such as imbalanced workload (a.k.a. skewed degree distribution)~\cite{powergraph,effective,pregelplus}, redundancies in communication~\cite{gps,xpregel,effective} and low convergence speed~\cite{thinkgraph,goffish,blogel}.
However, there remains one challenge: although the usefulness of these optimizations are well demonstrated in solving simple algorithms such as PageRank and single-source shortest path (SSSP)\footnote{\small PageRank and SSSP are basically a loop executing a simple computation kernel.}, it is, however, hard to combine them together to implement complex algorithms, where we may have to deal with multiple performance issues at the same time.

To see this challenge clearly,
let us consider the S-V algorithm~\cite{ShVi82,connectivity}, a known algorithm for computing connected components in undirected graph, which can be regarded as a distributed union-find algorithm~\cite{disjointset}.
Essentially, it is an iterative algorithm with two key operations --- pointer jumping and tree merging.
For the pointer jumping operation, the communication suffers from imbalanced workload~\cite{effective}, and in the meantime in the tree merging operation, the neighborhood communication (every vertex broadcasts a message to all of its own neighbors) could be potentially very heavy.
Although there are techniques in separate systems~\cite{gps,xpregel,effective} dealing with each case, there is no system capable of optimizing away both issues at the same time.
The main reason is Pregel's monolithic message mechanism.
When having multiple communication patterns in a single Pregel program, the system has no idea which message is for what purpose, thus it can do nothing for optimization.


In this paper, we propose a new approach to composing various optimizations together, by making use of the interface called {\em channel}~\cite{husky} as a replacement of Pregel's message passing mechanism.
Informally, a \text{channel} is responsible for sending or receiving messages of a certain pattern for some purpose (such as reading all neighbors' states, requesting data from some other vertex and so on).
And by slicing the messages by their purpose and organizing them in channels, we can characterize each channel by high-level communication patterns, identify the redundancies or potential performance issues, and then apply suitable optimizations to deal with the problems.

The technical contributions of this work can be summarized as follows.
\begin{itemize}[leftmargin=*]
\item First, we provide Pregel with a channel-based vertex-centric programming interface, which is intuitive in the sense that it is just a natural extension of Pregel's monolithic message mechanism.
To demonstrate the power of the channel interface, we implement three optimizations as special channels and show how they are easily composed to optimize complex algorithms such as the above S-V algorithm.


\item Second, we have fully implemented the system and the experiment results convincingly show the usefulness of our approach.
The channel interface itself contributes to a message reduction up to $82\%$ especially for complex algorithms, and the three optimized channels further improve the performance of the algorithms they are applicable (3.16x for PageRank, 3.29x for Pointer-Jumping and 5.02x for weakly connected components).
Specially, the composition of different optimizations makes the S-V algorithm 2.20x faster than the best implementation available now.
\end{itemize}

The rest of the paper is organized as follows.
\autoref{sec:background} briefly reviews the basic concepts of Pregel,
\autoref{sec:programming} introduces the programming interface of our channel-based system, as well as non-trivial examples showing how optimizations are composed in the same algorithm.
\autoref{sec:impl} presents the channel mechanism and the implementation of our optimizations as channels.
\autoref{sec:experiments} presents the experiment results, \autoref{sec:related} discusses the related work, and \autoref{sec:conclusion} concludes the paper.

\section{Background}
\label{sec:background}

In this section, we give a brief introduction to Pregel and illustrate the limitations of its message passing interface.

\subsection{Pregel}

A Pregel computation consists of a series of supersteps separated by global synchronization points.
In each superstep, the vertices compute in parallel executing the same user-defined function (usually the \texttt{compute()} method) that expresses the logic of a given algorithm.
A vertex can read the messages sent to it in the previous superstep, mutate its state, and send messages to its neighbors or any known vertex in the graph.
The termination of the algorithm is based on every vertex voting to halt.
Vertices can be reactivated externally by receiving messages.

Pregel provides the message passing interface for inter-vertex communication and aggregator for global communication.

\textbf{Message passing and the combiner.}
In Pregel, vertices communicate directly with each other by sending messages, where each message consists of a message value and a destination.
The combiner optimization~\cite{pregel} is applicable if the receiver only needs the aggregated result (like the sum, or the minimum) of all message values,
in which case the system is provided an associative binary function to combine messages for the same destination whenever possible.

\textbf{Aggregator.}
Aggregator is a useful interface for global communication, where each active vertex provides a value, and the system aggregates them to a final result using a user-specified operation and makes it available to all vertices in the next superstep.

\subsection{Problems in Pregel's Message Interface}

Pregel is designed to support iterative computations for graphs, and it is indeed suitable for algorithms like the PageRank or SSSP.
However, it is noteworthy that vertex-centric graph algorithms are in general complex.
Even for some fundamental problems like connected component (CC), strongly connected component (SCC) and minimum spanning forest (MSF), their efficient vertex-centric solutions require multiple computation phases, each having different communication patterns~\cite{connectivity,optimizing}.
For such complex algorithms, all the computation phases have to share Pregel's message passing interface, which causes the following problems:
\begin{itemize}[leftmargin=*]
 \item When different message types are needed in different computation phases, the Pregel's message interface has to be instantiated with a type that large enough to carry all those message values.
 \item Usually, we can no longer optimize any of the communication patterns in these computation phases, since the system cannot distinguish which message is to be optimized.
\end{itemize}

As mentioned before, these are the consequences of Pregel's monolithic message mechanism, which may not only increase the message size, but also prevent the possible optimizations to be applied.
We have detailed discussions using S-V in \autoref{sec:sv-algo} and evaluation results showing the message overhead (up to 5x) in \autoref{sec:eval-channel}.
It motivates us to design a better communication interface for Pregel to handle a wide range of complex vertex-centric graph algorithms.

\begin{table*}[t]
\centering
\caption{The APIs for standard channels.}
\label{tab:basic-apis}
\begin{tabular}{l|l|l}
\hline
  \multicolumn{2}{c|}{\textbf{Message-Passing Channels}}
& \textbf{Aggregator Channel} \\
\hline
  \inlineCPP{DirectMessage(Worker<VertexT> *w);}
& \inlineCPP{CombinedMessage(Worker<VertexT> *w,}
& \inlineCPP{Aggregator(Worker<VertexT> *w,} \\
 
& \inlineCPP{\ \ \ \ Combiner<ValT> c);}
& \inlineCPP{\ \ \ \ Combiner<ValT> c);} \\
\hline
  \inlineCPP{void send_message(KeyT dst, ValT m);}
& \inlineCPP{void send_message(KeyT dst, ValT m);}
& \inlineCPP{void add(ValT v);} \\
  \inlineCPP{MsgIterator<KeyT, ValT> \&get_iterator();}
& \inlineCPP{const ValT \&get_message();}
& \inlineCPP{const ValT \&result();} \\
\hline
\end{tabular}
\end{table*}

\begin{table*}[t]
\centering
\caption{The APIs for optimized channels.}
\label{tab:apis}
\begin{tabular}{l|l|l}
\hline
  \textbf{Scatter-Combine}
& \textbf{Request-Respond}
& \textbf{Propagation (Simplified)} \\
\hline
  \inlineCPP{ScatterCombine(Worker<VertexT> *w,}
& \inlineCPP{RequestRespond(Worker<VertexT> *w,}
& \inlineCPP{Propagation(Worker<VertexT> *w,} \\
  \inlineCPP{\ \ \ \ Combiner<ValT> c);}
& \inlineCPP{\ \ \ \ function<RespT(VertexT)> f);}
& \inlineCPP{\ \ \ \ Combiner<ValT> c);} \\
\hline
  \inlineCPP{void add_edge(KeyT dst);}
& 
& \inlineCPP{void add_edge(KeyT dst);} \\
\hline
  \inlineCPP{void set_message(ValT m);}
& \inlineCPP{void add_request(KeyT dst);}
& \inlineCPP{void set_value(ValT m);} \\
  \inlineCPP{const ValT \&get_message();}
& \inlineCPP{const RespT \&get_respond();}
& \inlineCPP{const ValT \&get_value();} \\
\hline
\end{tabular}
\end{table*}

\section{Programming with Channels}
\label{sec:programming}

The channel mechanism is designed to help users better organize the communications in vertex-centric graph algorithms.
Concretely speaking, the channels are message containers equipped with a set of methods for sending/receiving messages or supporting a specific communication pattern (see \autoref{tab:basic-apis} and \autoref{tab:apis} for the standard and optimized channels; the details are in \autoref{sec:opt-channels}).
In this section, we first introduce the programming interface using the PageRank example, then we show how different optimizations can be easily composed via channels in a more complex algorithm called the S-V~\cite{ShVi82}.

\subsection{A Standard PageRank Implementation Using Channels}

Writing a vertex-centric algorithm in our system using the standard channels is rather straightforward for a Pregel programmer.
We present a PageRank Implementation in \autoref{fig:pagerank-channel}, which is basically obtained from a Pregel program by replacing the sending/reading of messages by a user-defined message channel's send/receive methods.

In the first 30 supersteps, each vertex sends along outgoing edges (if exists) its tentative PageRank divided by the number of outgoing edges (lines 21--25), over a user-defined message channel \texttt{msg}.
This channel is an instance of \texttt{CombinedMessage}, so a combiner is provided to the channel's constructor (line 9).
In the next superstep, a vertex gets the sum of the values arriving on this channel (lines 18) and calculate a new PageRank.
To avoid PageRank lost in dead ends (vertices without outgoing edges), we need a \textit{sink} node to collect the PageRank from those dead ends and redistribute it to all nodes, which is implemented by an aggregator \texttt{agg}.
Users explicitly add the PageRank to the aggregator (line 27) and reads the aggregated result in the next superstep (line 16).

\begin{figure}[t]
\begin{lstlisting}[basicstyle=\scriptsize\ttfamily,language=c++,morekeywords={override},
numbers=left,stepnumber=1,xleftmargin=2.5em,commentstyle=\color{mygreen}]
using VertexT = Vertex<int, PRValue>;
auto c = make_combiner(c_sum, 0.0); // a combiner
class PageRankWorker : public Worker<VertexT> {
private:
   // two channels are defined here
   CombinedMessage<VertexT, double> msg;
   Aggregator<VertexT, double> agg;
public:
   PageRankWorker():msg(this, c), agg(this, c) {}

   void compute(VertexT &v) override {
      if (step_num() == 1) {
         value().PageRank = 1.0 / get_vnum();
      } else {
         // s: the pagerank of the "sink node"
         double s = agg.result() / get_vnum();
         value().PageRank = 0.15 / get_vnum()
               + 0.85 * (msg.get_message() + s);
      }
      if (step_num() < 31) {
         int numEdges = value().Edges.size();
         if (numEdges > 0) {
            double msg = value().PageRank / numEdges;
            for (int e : value().Edges)
               msg.send_message(e, msg);
         } else
            agg.add(value().PageRank);
      } else
         vote_to_halt();
   }
};
\end{lstlisting}
\caption{PageRank implementation using channels.}
\label{fig:pagerank-channel}
\end{figure}

\subsection{Channels and Optimizations}

In our channel-based system, we offer a set of optimizations as special channels (in \autoref{tab:apis}), which can be regarded as more efficient implementations (compared to the standard message passing channels) of several communication patterns.

Consider the PageRank example (\autoref{fig:pagerank-channel}), by simply changing one line in the channel definition, we can enable the scatter-combine optimization that handles the ``static messaging pattern'':
\begin{lstlisting}[style=highlight,firstnumber=5]
   // change to a scatter-combine channel
   ScatterCombine<VertexT, double> msg;
\end{lstlisting}
Then, in the \texttt{compute()} function, the programmer of course needs to initialize the scatter-combine channel (by invoking \texttt{add\_edge()}) and switches to its dedicated interface, but the total changes are just five lines of code.
Our experiments (\autoref{sec:eval-sc}) show that, by switching to the scatter-combine channel, the PageRank immediately gets 2x--5x faster, and all the programmer need to understand is the high-level abstraction of each channel.

\subsection{Composition of Channels}
\label{sec:sv-algo}


In this part, we use a more complicated example called the Shiloach-Vishkin (S-V) algorithm~\cite{ShVi82} to show that,
users can easily combine different optimizations (channels) to handle multiple performance issues in the same program.

\subsubsection{The S-V Algorithm}


The S-V algorithm is in general an adoption of the classic union-find algorithm~\cite{disjointset} to the distributed setting, which finds the connected components in undirected graphs with $n$ vertices in $O(\log n)$ supersteps.
A distributed tree structured (called the disjoint-set) is maintained by each vertex holding a pointer to either some other vertex in the same connected component or to itself if it is a root.
We henceforth use $D[u]$ to represent this pointer for vertex $u$.
Following is the high-level description of the S-V algorithm using a domain-specific language called Palgol~\cite{palgol}, and it compiles to Pregel code.

\begin{lstlisting}[basicstyle=\scriptsize\ttfamily,language=palgol,
numbers=left,stepnumber=1,xleftmargin=2.5em,commentstyle=\color{mygreen},numberfirstline=false]
// initially suppose we have D[u] = u for every u
do
  // enter vertex-centric mode
  for u in V
    // whether u's parent is a root vertex
    if (D[D[u]] == D[u])
      // iterate over neighbors (D[e]: neighbor's pointer)
      let t = minimum [ D[e] | e <- Nbr[u] ]
      if (t < D[u])
        // modify the D field of u's parent D[u]
        remote D[D[u]] <?= t
    else
      // the pointer jumping (path compression)
      D[u] := D[D[u]]
  end
until fix[D] // until D stabilizes for every u
\end{lstlisting}

Starting from $n$ root nodes, the S-V algorithm iteratively merges the trees together if crossing edges are detected.
In a vertex-centric way, every vertex~$u$ performs one of the following operations depending on whether its parent $D[u]$ is a root vertex:
\begin{itemize}[leftmargin=*]
 \item \textbf{Tree merging (lines 7--11).}
  If $D[u]$ is a root vertex, $u$ sends the smallest one of its neighbors' pointer (to which we give a name $t$) to the root $D[u]$ and later the root points to the minimum $t$ it receives (to guarantee the correctness of the algorithm).
 \item \textbf{Pointer jumping (line 14).}
  If $D[u]$ is not a root vertex, $u$ modifies its pointer to its ``grandfather'' ($D[u]$'s current pointer), which halves the distance to the current root.
\end{itemize}
The algorithm terminates when all vertices' pointers do not change after an iteration.
Readers interested in the correctness of this algorithm can be found in the original paper~\cite{connectivity} for more details.

\subsubsection{Choices of Channels}

In the S-V algorithm, three major performance issues are identified below by analyzing the communication patterns in the algorithm.
\begin{itemize}[leftmargin=*]
 \item The load balance issue in testing whether $D[u]$ is a root vertex or not for every $u$.
  The standard implementation is to let each $u$ send a request to its current parent $D[u]$, then the reply message is the parent's pointer.
  A few vertices with very large degree may block the reply phase.
 \item The heavy neighborhood communication in calculating the minimum parent ID of the neighboring vertices, where all vertices need to send a message to each neighbor, regardless of the vertices' local state.
 \item The congestion issue in the modification of parent's pointer, due to the existence of high-degree vertices.
\end{itemize}

Fortunately, our system has already provided all the solutions to these issues, while users just need to choose the proper channels for each kind of communication pattern.
The load balance issue can be avoided by the request-respond channel, the heavy neighborhood communication is optimized by the scatter-combine channel, and a combined-message channel solves the congestion issue.

\section{Channel Implementation}
\label{sec:impl}

In this section, we present the design of our channel mechanism, and demonstrate how three interesting channels can be implemented for capturing three optimizations.

\subsection{Overview}

\begin{figure}[t]
 \centering
 \includegraphics[width=0.5\textwidth]{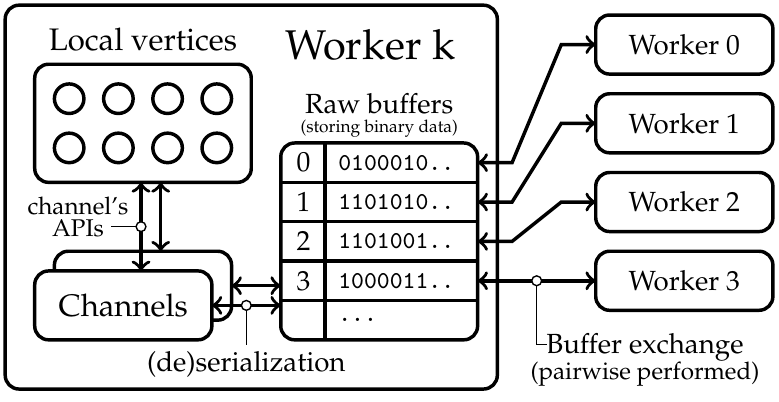}
 \caption{The architecture of our channel-based system.}
 \label{fig:channels}
\end{figure}

\autoref{fig:channels} shows the architecture of our channel-based system.
Worker is the basic computing unit in our system.
When launching a graph processing task, multiple instances of workers are created, each holding a disjoint portion of the graph (a subset of vertices along with their states and adjacent lists).
Workers share no memory but can communicate with each other.
Such big picture is common in all Pregel systems, but the main difference is the hierarchy of the components inside the worker.

In our system, channels form an independent layer inside the worker between the vertices and the raw buffers.
Each worker has $M-1$ buffers (where $M$ is the number of workers launched by the user) for storing binary message data for each other worker, then the channels can read or write these buffers in its own address space.
Each channel independently implements a communication pattern (like messages passing or aggregator) and provides its own interfaces for the vertices, and users choose the proper channels to implement a graph algorithm according to the communication patterns it has.

\begin{figure}[t]
\begin{lstlisting}[basicstyle=\scriptsize\ttfamily,language=c++,morekeywords={override},
numbers=left,stepnumber=1,xleftmargin=2.5em,commentstyle=\color{mygreen}]
class Channel {
public:
   // initialization function
   virtual void initialize() {};
   // paired (de)serialization functions
   virtual void serialize(Buffer &buff) = 0;
   virtual void deserialize(Buffer &buff) = 0;
   // return true for additional buffer exchange
   virtual bool again() { return false; };
};
\end{lstlisting}
\caption{The core functions of the base class \texttt{Channel}.}
\label{fig:channel-class}
\end{figure}

\subsection{Channels}


The channel is designed for allowing experts to implement new optimizations with ease.
\autoref{fig:channel-class} shows the base class \texttt{Channel} and its core functions: \texttt{initialize()}, \texttt{serialize()} for writing data to worker's raw buffer, \texttt{deserialize()} for reading data from worker's raw buffer (after the buffer exchange) and \texttt{again()} for supporting multiple rounds of communication.
All the channels in our paper are implemented as derived classes of \texttt{Channel} with proper implementations of these four functions (in particular \texttt{serialize()} and \texttt{deserialize()}).

To clearly see how the workers and channels cooperate with each other, we present the computation logic of the worker in \autoref{fig:channel-logic}.
The worker's computation is organized as synchronized supersteps.
In each superstep, the worker first calls the \texttt{compute()} on every vertex, then it performs several rounds of buffer exchanges.
In each round, the system invokes the active channels' \texttt{serialize()} and \texttt{deserialize()} methods to exchange data between the channels and the buffers.
All channels are set to active at the beginning, but they can deactivate themselves by returning \texttt{false} in the \texttt{again()} function.
Channels' \texttt{initialize()} is invoked at the beginning of the computation.
While not explicitly presented in the code, the channels can activate vertices through the \texttt{Worker}'s interface by providing the vertex's ID or local index.
That is how our system simulates the voting-to-halt mechanism of Pregel.

\begin{figure}[t]
\begin{lstlisting}[style=highlight,
morekeywords={foreach,end_for,end_while}]
load_graph()
foreach channel c do c.@initialize@()
foreach vertex v do v.set_active(true)
while (active vertex exists) // a superstep
   foreach active vertex v do this.compute(v)
   foreach channel c do c.set_active(true)
   while (active channel exists)
      foreach active channel c do c.@serialize@()
      buffer_exchange()
      foreach active channel c do
         c.@deserialize@()
         c.set_active(c.@again@())
      end_for
   end_while
end_while
dump_graph()
\end{lstlisting}
\caption{The computation logic of the worker for illustrating the channel mechanism.}
\label{fig:channel-logic}
\end{figure}

\subsection{Optimized Channels}
\label{sec:opt-channels}

As the last part of this section, we give a brief introduction of the three optimized channels currently provided in our library, which target three important performance issues in vertex-centric graph processing.

\subsubsection{Scatter-Combine Channel}
\label{sec:scatter-combine}

The scatter-combine abstraction is a common high-level pattern appeared in many single-phase algorithms such as PageRank, single-source shortest path (SSSP) and connected component (CC).
The communication in this model is captured by a \texttt{scatter()} function on each vertex to send the same value to all neighbors, and a \texttt{combine()} function to combine the messages for each receiver.
We focus on a special case where \textit{every} vertex needs to send a value to all of its neighbors\footnote{\small In some algorithms like SSSP or WCC, only active vertices need to send messages, which is not the case we are targeting here.} regardless of its local state.
An iterative algorithm having such static messaging pattern will waste time repeating the same message dispatching procedure, while a proper pre-processing can greatly reduce the computation time as well as the message size.


\begin{figure}[t]
 \centering
 \includegraphics[width=0.46\textwidth]{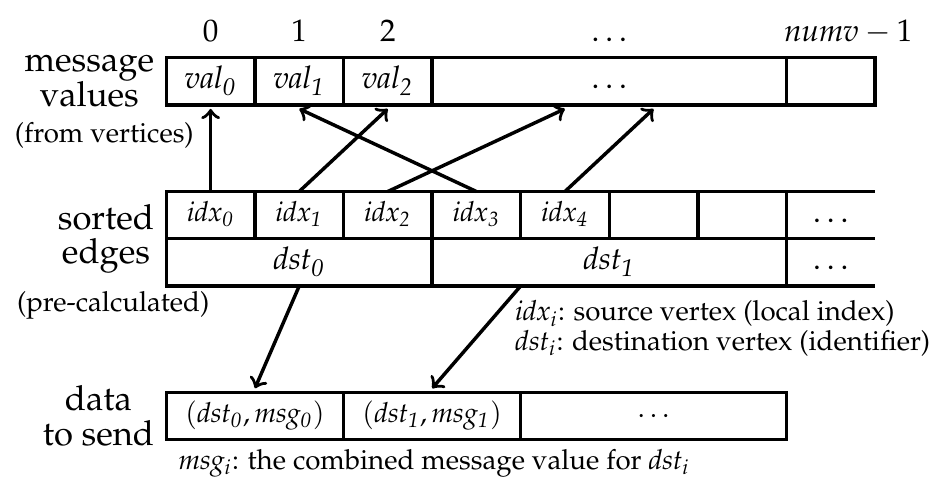}
 \caption{The execution logic for the scatter-combine channel.}
 \label{fig:scatter-impl}
 \vspace{-2pt}
\end{figure}

\autoref{fig:scatter-impl} demonstrates the computation logic of the scatter-combine channel.
Suppose the vertices on an worker is indexed by \texttt{0..numv-1}, then each local edge is a pair $(idx, dst)$ where $idx$ refers to a local vertex and the $dst$ can be an arbitrary vertex in the graph.
We sort the edges by $dst$ in advance, then by scanning the array of the sorted edge list once, we can quickly calculate for each destination a combined message value.
This is much cheaper than the normal message routine which typically requires hashing or sorting.



The APIs for the scatter-combine channel are presented in the first column of \autoref{tab:apis}.
Users need to initialize the channel by adding the outgoing edges of each vertex through the \texttt{add\_edge()} function before the first message sending occurs in the execution.
Then, every vertex emits an initial messages using the \texttt{send\_message()} interface and the combined messages for each vertex can be obtained by the \texttt{get\_message()} method in the next superstep.

\subsubsection{Request-Respond Channel}
\label{sec:request-respond}

This is a communication pattern where two rounds of message passing (say the request phase and respond phase) together form a conversation to let every vertex request an attribute of another vertex.
Typically, such computation contains vertices with high degree which causes imbalanced workload in the respond phase, and the solution is to merge the requests of the same destination on each worker.
More details can be found in the original paper~\cite{effective}.

\begin{figure}[t]
 \centering
 \includegraphics[width=0.46\textwidth]{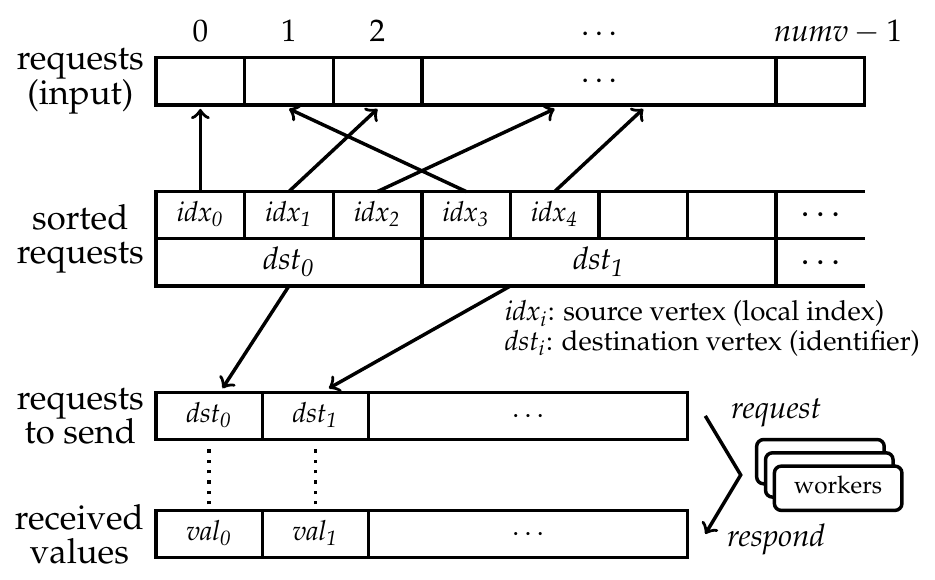}
 \caption{The execution logic for the request-respond channel.}
 \label{fig:reqresp-impl}
 \vspace{-2pt}
\end{figure}

Our implementation of this optimization is illustrated in \autoref{fig:reqresp-impl}.
A request is a pair $(idx, dst)$ where $idx$ refers to a local requester and the $dst$ can be an arbitrary vertex in the graph.
The worker sorts the requests by $dst$ and sends exactly one message containing the worker ID to each of the unique destinations.
When receiving the response values, the worker performs another scan to the sorted requests, which is sufficient to reply to all the requesters.

The middle column of \autoref{tab:apis} shows the APIs of the request-respond channel.
When creating the channel, users need to provide a function that generates a response value from a vertex value.
The whole procedure is implemented in an implicit style; A vertex invokes \texttt{add\_request()} with the destination vertex ID;
all the requests are delivered after the request phase, and the vertices receiving any request will be automatically involved, and a response value is produced by the user-provided function.

\subsubsection{Propagation Channel}
\label{sec:propagation}

The last optimized channel is to speedup the convergence for a class of propagation-based algorithms.
In these algorithms, typically, some vertices emit the initial labels, and in each of the following supersteps, vertices receiving the labels will perform some computation and may further propagate a new label to their outgoing neighbors.
Since the propagation is between neighbors, such algorithms converge very slowly on graphs with large diameters.

The design of this channel is inspired by two existing techniques for improving the convergence speed.
First, the GAS model~\cite{powergraph} with an asynchronous execution mode can perform the crucial updates as early as possible without waiting for the global synchronization.
Although this implementation is not feasible in our synchronous system, the high-level abstraction is suitable for describing such kind of computation.
Second, the block-centric computation model~\cite{thinkgraph,blogel,goffish} is an extension of Pregel which opens the partition to users, so that users can choose a suitable partition method and implement a block-level computation to perform the label propagation within a connected subgraph.

Our propagation channel combines the advantages of these two techniques: it provides a simplified GAS model which naturally describes such propagation-based computation, and its implementation works in a similar way as a block-level program to accelerate the label propagation.
Therefore, users allocate a channel to obtain a performance gain without additional efforts on writing the block-level program.

\begin{figure}[t]
\centering
\includegraphics[width=0.48\textwidth]{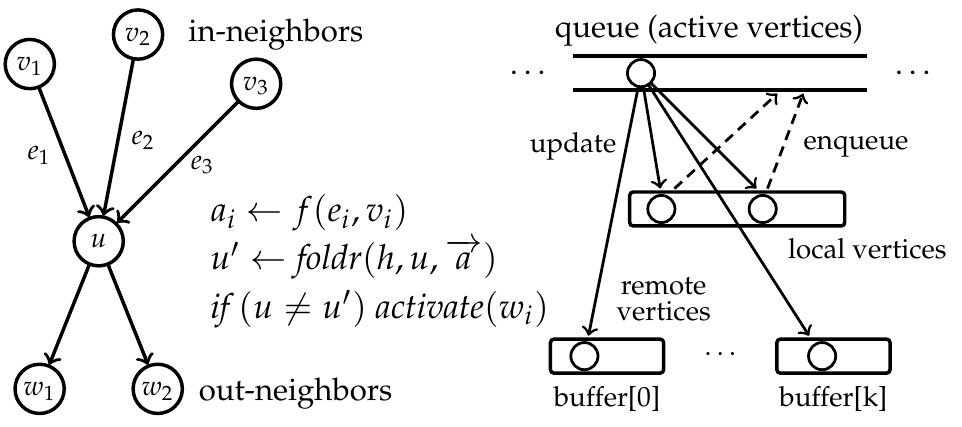}
\caption{The propagation channel's high-level model and computation logic.}
\label{fig:propagation-model}
\end{figure}

\autoref{fig:propagation-model} describes the high-level model for the propagation channel as well as the execution logic in our implementation.
Initially, each vertex is associated with a value and is set to active.
Whenever having an active vertex $u$ in the graph, it reads each incoming neighbors and the corresponding edges (if exists), and calculate a value $a_i$ by a user-provided function $f$.
Then, a combiner $h$ updates the original vertex value $u$ by each neighbor's $a_i$ and returns a new vertex value $u'$.
If the new value $u'$ is different from the original value $u$, we activate all outgoing neighbors of $u$ to propagate the update, and finally $u$ is deactivated after being processed.
The computation stops when all the vertices are inactive.
Note that we require $h$ to be commutative, so that the order of combining $a_i$ does not affect the result.
Moreover, when any of the incoming neighbors of $u$ is modified, $u$ needs to read the modified vertex to update its own value, instead of recomputing the $\mathit{foldr}$ by reading all its incoming neighbors' values.

This computation model is implemented by each worker performing a BFS-like traversal on the subgraph it holds.
Starting from the initial setting, each worker propagates the values along the edges as far as possible.
It updates the local vertices directly, but records the changes on remote vertices as messages.
The buffer exchange is performed after no local updates is viable on any worker.
After the remote updates triggered by messages, a new round of local propagation is performed.
It terminates when all vertices have converged.

The last column of \autoref{tab:apis} shows the APIs of a simplified propagation channel without considering the edge weights (for saving space), so users provide a combiner to calculate the new vertex value.
Each vertex adds its adjacent list to the channel via \texttt{add\_edge()} and sets the initial value by \texttt{set\_value()}, and in the next superstep, a vertex invokes \texttt{get\_value()} to get the final value after the propagation converges.
To make the best use of the propagation channel, users should preprocess the graph by tagging a partition ID to the vertex IDs.

\section{Experiments}
\label{sec:experiments}

The experiments are conducted on an Amazon EC2 cluster of 8 nodes (with instance type m4.xlarge), each having 4 vCPUs and 16G memory.
The connectivity between any pair of nodes in the cluster is 750Mbps.
The datasets are listed in \autoref{tab:datasets} including both real-world graphs (Wikipedia\footnote{\small\url{http://konect.uni-koblenz.de/networks/dbpedia-link}}, Facebook\footnote{\small\url{http://archive.is/tVl1G}}, WebUK\footnote{\small\url{http://networkrepository.com/web_uk_2005_all.php}}, Twitter\footnote{\small\url{http://konect.uni-koblenz.de/networks/twitter_mpi}}(\textit{undirected}) and USA road network\footnote{\small\url{http://www.dis.uniroma1.it/challenge9/download.shtml}}) and synthesized graphs (a chain, a random tree and a weighted power-law graph using R-MAT~\cite{rmat}).

We select six representative algorithms in our evaluation, including PageRank (PR), Pointer-Jumping (PJ), Weakly Connected Component (WCC), S-V algorithm (S-V), Strongly Connected Component (SCC) and Minimum Spanning Forest (MSF).
For comparison, we also present the results of our best-effort implementations in \PP{} \cite{pregelplus} and Blogel~\cite{blogel}.
Both of them are typical Pregel implementations, where \PP{} supports the request-respond paradigm and mirroring technique in two special modes (\textit{reqresp} mode and \textit{ghost} mode respectively) and Blogel supports the block-centric computation.
All of these systems mentioned above as well as our channel-based system are implemented in \CC{} on top of the Hadoop Distributed File System (HDFS).
The source code of our system can be accessed at \url{https://bitbucket.org/zyz915/pregel-channel}.

\begin{table}[t]
\centering
\caption{Datasets used in our evaluation}
\label{tab:datasets}
\begin{tabular}{c|c|c|c|c}
\hline
\textbf{Dataset} & \textbf{Type} & $|V|$ & $|E|$ & avg. Deg \\
\hline\hline
Wikipedia & \multirow{2}{*}{directed} & 18.27M & 172.31M & 9.43 \\
\cline{1-1}\cline{3-5}
WebUK & & 39.45M & 936.36M & 23.73 \\
\hline\hline
Facebook & \multirow{2}{*}{undirected} & 59.22M & 185.04M & 3.12 \\
\cline{1-1}\cline{3-5}
Twitter & & 41.65M & 2.94B & 70.51 \\
\hline\hline
Tree$^{*}$ & \multirow{2}{*}{rooted tree} & 100M & 100M & 1.00 \\
\cline{1-1}\cline{3-5}
Chain$^{*}$ & &  100M & 100M & 1.00 \\
\hline\hline
USA Road & undirected & 23.95M & 57.71M & 2.41 \\
\cline{1-1}\cline{3-5}
RMAT24$^{*}$ & \& weighted & 16.78M & 268.44M & 16.00 \\
\hline
\multicolumn{5}{c}{\hbox{datasets marked with $*$ are synthetic.}}
\end{tabular}
\end{table}

\subsection{The Channel Mechanism}
\label{sec:eval-channel}

First, we evaluate the standard channels (the message passing channels and aggregator) in our system.
Basically, rewriting a Pregel program into a channel-based version is just about replacing the matched send-receive pairs into the same channel's send/receive function.
The message is chosen as small as possible, and we always use a combiner if applicable.
We compare both implementations to see whether there is any overhead or benefits introduced by our channel mechanism.

The experiment results are presented in \autoref{tab:eval-cmp}, where a straightforward rewriting achieves a speedup ranging from 1.08x to 2.64x among all the five algorithms on those datasets.
For MSF, SCC and S-V, we also observe a significant reduction on message size ranging from $23\%$ to $82\%$.

\textbf{Analysis.}
The channel mechanism itself can improve the performance, due to the following two reasons.
First, our system allows users to specify a combiner to a channel whenever applicable, 
while in Pregel, we can specify a global combiner only when all the messages in the algorithm can use that combiner.
This difference makes our S-V and SCC more message-efficient.
where the inapplicability of combiner in \PP{} causes a 5.52x message size on Twitter and 1.55x message size on Facebook.

Second, our channel-based system allows users to choose different message types for different channels, while in \PP{}, a global message type is chosen to serve all communication in the program.
Then, the MSF (we refer to a particular version here~\cite{boruvka}) is a typical example that uses heterogeneous messages in different phases of the algorithm.
The largest message type is a 4-tuple of integer values for storing an edge, but the smallest one is just an \texttt{int}.

The channel interface does introduce some overhead, which causes our SCC implementation slightly worse than the one in Pregel+.
However, such overhead is only visible when the message synchronization dominates the execution time.
For SCC, it terminates in 1247 supersteps and in most supersteps the number of active vertices is extremely few (less than $0.01\%$).

For the rest algorithms, there is no significant difference when implemented in two systems.
Still, our system reduces the runtime of PR and WCC by up to $13\%$ and $8\%$ (using the \texttt{CombinedMessage} class), and for PJ (using the \texttt{DirectMessage} class) the number is $45\%$.
We believe that the improvement is due to the choice of message interface (in particular the message iterator in \texttt{DirectMessage} instead of nested \CC{} vectors in \PP{}).
Nevertheless, we show that our system implementation is reasonably efficient.

\begin{table}[t]
\centering
\caption{Comparison of the basic implementation of graph algorithms in Pregel+ and channel-based system.}
\label{tab:eval-cmp}
\begin{tabular}{c|c|c|c|c}
\hline
\multirow{2}{*}{PR} & \multicolumn{2}{|c}{WebUK} & \multicolumn{2}{|c}{Wikipedia} \\
\cline{2-5}
 & pregel & channel & pregel & channel \\
\hline
runtime (s) & 212.24 & 205.80 & 47.32 & 40.36 \\
\hline
message (GB) & 63.23 & 63.23 & 14.02 & 14.02 \\
\hline
\hline
\multirow{2}{*}{WCC} & \multicolumn{2}{|c}{Wikipedia} & \multicolumn{2}{|c}{Wikipedia (P)} \\
\cline{2-5}
 & pregel & channel & pregel & channel \\
\hline
runtime (s) & 16.96 & 15.67 & 15.31 & 15.85 \\
\hline
message (GB) & 2.85 & 2.85 & 0.49 & 0.49 \\
\hline
\hline
\multirow{2}{*}{PJ} & \multicolumn{2}{|c}{Chain} & \multicolumn{2}{|c}{Tree} \\
\cline{2-5}
 & pregel & channel & pregel & channel \\
\hline
runtime (s) & 111.54 & 69.63 & 36.25 & 19.94 \\
\hline
message (GB) & 39.99 & 39.99 & 8.56 & 8.56 \\
\hline
\hline
\multirow{2}{*}{S-V} & \multicolumn{2}{|c}{Facebook} & \multicolumn{2}{|c}{Twitter} \\
\cline{2-5}
 & pregel & channel & pregel & channel \\
\hline
runtime (s) & 49.74 & 37.92 & 382.60 & 144.99 \\
\hline
message (GB) & 16.41 & 11.46 & 112.21 & 20.32 \\
\hline
\hline
\multirow{2}{*}{MSF} & \multicolumn{2}{|c}{USA} & \multicolumn{2}{|c}{RMAT24} \\
\cline{2-5}
 & pregel & channel & pregel & channel \\
\hline
runtime (s) & 27.05 & 16.13 & 50.56 & 45.94 \\
\hline
message (GB) & 8.67 & 4.86 & 14.80 & 12.91 \\
\hline
\hline
\multirow{2}{*}{SCC} & \multicolumn{2}{|c}{Wikipedia} & \multicolumn{2}{|c}{Wikipedia (P)} \\
\cline{2-5}
 & pregel & channel & pregel & channel \\
\hline
runtime (s) & 52.15 & 61.89 & 50.51 & 67.84 \\
\hline
message (GB) & 9.85 & 4.98 & 2.70 & 1.29 \\
\hline
\end{tabular}
\end{table}

\subsection{Effectiveness of Optimized Channels}

Here, we evaluate the efficiency of our optimized channels against the message passing channels using the applications that each kind of channel is applicable.
In this part, we choose rather simple algorithms, so that we can clearly see how optimized channels can improve the performance in different scenarios.

\subsubsection{Scatter-Combine Channel}
\label{sec:eval-sc}

PageRank is a typical graph algorithm that can use the scatter-combine algorithm.
We test \PP{}'s basic implementation, \PP{}'s ghost mode (a.k.a. the mirroring technique~\cite{effective}), the standard channel version (\autoref{fig:pagerank-channel}) and the scatter-combine channel version.
For \PP{}'s mirroring technique, we set the threshold to 16 in all cases.

The experiment results are presented in the upper part of \autoref{tab:results}.
The basic mode of \PP{} and our standard version are close in both execution time and message size, while the scatter-combine channel achieves a speedup ranging from 3.03x to 3.16x and reduces roughly one third of the message size.
\PP{}'s ghost mode use less messages, but the execution time (including the preprocessing time) is not reduced significantly on Wikipedia and is even worse on WebUK.

\textbf{Analysis.}
The improvement on execution time clearly shows the effectiveness of the scatter-combine channel.
As explained in \autoref{sec:scatter-combine}, it can generate the combined messages by a linear scan of the edges, while \PP{}'s basic mode and the \texttt{CombinedMessages} have to use hash table or sorting to handle the general case.
The reduction on total message size is explained by the removal of redundant transmission of vertices' identifiers.

All these three programs use the \textit{receiver-centric} message combining, while \PP{}'s mirroring technique uses the \textit{sender-centric} message combining to further reduce the messages in transmission, where a high-degree vertex only sends at most one message to each other worker, instead of one message to each neighbor.
However, such method is computational intensive and the overall computational cost is still high due to the involvement of hash table.
We show that the computational cost in message processing is a major problem in some algorithms, and our scatter-combine achieves better performance than existing approaches.

\subsubsection{Request-Respond Channel}
\label{sec:eval-rr}

This optimization is originally introduced in \PP{}'s \textit{reqresp} mode and it is interesting to see which implementation is faster.
A representative application is the pointer-jumping algorithm.
Given a (forest of) rooted tree, each vertex initially knows its parent and tries to find the root of the tree it belongs to.
The algorithm is actually part of the S-V (\autoref{sec:sv-algo}).
We can consider it as a minimum example that uses the request-respond paradigm.

We test pointer-jumping over two types of graphs, one is a randomly generated tree, and the other is a chain.
Vertices are randomly assigned to workers.

\begin{table}[t]
\caption{Experiment results for each specialized channel.}
\label{tab:results}
\begin{tabular}{l|c|c|c|c}
\hline
\multicolumn{5}{c}{Scatter-Combine channel using PR} \\
\hline
\multicolumn{1}{c}{\multirow{2}{*}{Program}} & \multicolumn{2}{|c}{Wikipedia} & \multicolumn{2}{|c}{WebUK} \\
\cline{2-5}
& \small runtime & \small message & \small runtime & \small message \\
\hline
pregel+(basic) & 47.32 & 14.02  & 212.24 & 63.23  \\
\hline
pregel+(ghost) & 45.55 & 4.70  & 246.41 & 23.69  \\
\hline
channel (basic) & 40.36 & 14.02  & 205.80 & 63.23  \\
\hline
channel (scatter) & 15.58 & 9.50  & 67.00 & 42.86  \\
\hline
\hline
\multicolumn{5}{c}{Request-Respond channel using PJ} \\
\hline
\multicolumn{1}{c}{\multirow{2}{*}{Program}} & \multicolumn{2}{|c}{Tree} & \multicolumn{2}{|c}{Chain} \\
\cline{2-5}
& \small runtime & \small message & \small runtime & \small message \\
\hline
pregel+(basic) & 36.25 & 8.56  & 111.54 & 39.99  \\
\hline
pregel+(reqresp) & 54.37 & 2.62  & 676.19 & 28.87  \\
\hline
channel (basic) & 19.94 & 8.56  & 69.63 & 39.99  \\
\hline
channel (reqresp) & 11.03 & 1.75  & 74.10 & 19.24  \\
\hline
\hline
\multicolumn{5}{c}{Propagation channel using WCC} \\
\hline
\multicolumn{1}{c}{\multirow{2}{*}{Program}} & \multicolumn{2}{|c}{Wikipedia} & \multicolumn{2}{|c}{Wikipedia (P)} \\
\cline{2-5}
& \small runtime & \small message & \small runtime & \small message \\
\hline
pregel+(basic) & 16.96 & 2.85  & 15.31 & 0.49  \\
\hline
blogel & 20.39 & 1.11  & 5.10 & 0.11  \\
\hline
channel (basic) & 15.67 & 2.85  & 15.85 & 0.49  \\
\hline
channel (prop.) & 8.64 & 1.66  & 3.05 & 0.17  \\
\hline
\end{tabular}
\end{table}

The middle part of \autoref{tab:results} summarizes the results on these two types of graphs.
Without the request-respond optimization, the standard implementations in the two systems use exactly the same amount of messages, but ours runs 1.81x faster on a chain and 1.60x faster on a randomly generated tree.
Contrary to expectations, \PP{}'s \textit{reqresp} mode makes the program slower than its ordinary version, although the message size indeed decreases.
Using the same idea for optimization, our request-respond channel runs faster on a randomly generated tree, and is as good as an ordinary implementation when tree degrades to a chain.
Compared to \PP{}'s \textit{reqresp}, our request-respond channel reduces the message size is constantly $33\%$ less, and a performance gain up to 6.06x is observed on the Chain.

\textbf{Analysis.}
Although sharing the same idea, the implementations of the request-respond paradigm in our system and \PP{} are slightly different, which we believe is the main reason that makes our implementation better in both runtime and message size.
One particular trick in our implementation is that, when a worker sends the requests to another worker, it sends a list of vertex IDs, and the receiver sends back a list of values in exactly the same order.
However, in \PP{}, the receiver replies a pair of vertex ID and a value for each request, so that the message size increases. 

We also observe that, in real algorithms like S-V (\autoref{sec:sv-algo}), we are actually dealing with a dynamic forest, where the finding of the root vertex root is fused with the tree merging.
In this special case, \PP{}'s \textit{reqresp} mode can still make an improvement (see \autoref{tab:sv-results}).
Nevertheless, we verify that our implementation of the request-respond technique is reasonably effective, and is faster than the one in \PP{}.

\subsubsection{Propagation Channel}

We consider the HCC algorithm~\cite{pegasus} as a suitable example for using this optimization, which finds the weakly connected component (WCC) of a directed graph.
In this experiment, we present both the results on the original Wikipedia graph and the partitioned graph by METIS~\cite{metis}.
We also add the Blogel version here since the block-centric model is applicable~\cite{blogel}.
We choose METIS since it requires no additional knowledge of the graph.

The experiment results are presented in the bottom part of \autoref{tab:results}.
First, the \PP{} program and a standard channel version in our system are very close in both execution time (ours is $8\%$ faster) and message size (the same).
The block-centric version in Blogel works slightly worse on the original graph, but achieves roughly 3x faster when the input graph is properly partitioned.
Our propagation channel version works consistently better than all other implementations in terms of execution time on both graphs (1.67x faster than Blogel).
The number of messages used in the propagation channel version is the same as the Blogel version, but the message size in Blogel is $33\%$ less due to its special treatment of partition information.
Nevertheless, running WCC on partitioned graph is not message intensive.

\textbf{Analysis.}
A partitioner reduces the communication cost between the workers, but for the standard WCCs (program 1 and 3), it still takes a large number of supersteps to converge, so the execution time is not reduced.
Both of Blogel and our propagation channel use a block-level program to speedup the convergence and our system outperforms Blogel slightly.

It is also noteworthy that, WCC is quite simple that only needs around 10 lines of code for the \texttt{compute()} function.
Using the propagation channel in our system does not increase this number, while the Blogel version requires users to additionally write a block-level computation of more than 100 lines of code\footnote{\small\url{http://www.cse.cuhk.edu.hk/blogel/code/apps/block/hashmin/block.zip}}.
It is clear that our system achieves both conciseness and efficiency compared to the block-centric model.

\subsection{Combination of Channels}
\label{sec:eval-comb}

\begin{table}[t]
\caption{Experiment results of the S-V implementations using different combinations of channels.}
\label{tab:sv-results}
\begin{tabular}{l|c|c|c|c}
\hline
\multicolumn{1}{c}{\multirow{2}{*}{Program}} & \multicolumn{2}{|c}{Facebook} & \multicolumn{2}{|c}{Twitter} \\
\cline{2-5}
& \small runtime & \small message & \small runtime & \small message \\
\hline\hline
1-pregel+(reqresp) & 35.67 & 6.33  & 182.93 & 19.66  \\
\hline
2-channel (basic) & 37.92 & 11.46  & 144.99 & 20.32  \\
\hline
3-channel (reqresp) & 26.83 & 5.45  & 138.44 & 16.76  \\
\hline
4-channel (scatter) & 33.21 & 9.09  & 87.52 & 13.34  \\
\hline
5-channel (both) & 22.29 & 3.08  & 79.76 & 9.78  \\
\hline
\end{tabular}
\vspace{-2.5ex}
\end{table}

In this part, we verify the multiple performance issues in the S-V (see discussions in \autoref{sec:sv-algo}) by trying different combination of channels in our system.
We show that a combination of properly chosen channels can finally lead to much better performance.
To cover all the special channels we have,
we also present the experiment results of the Min-Label algorithm~\cite{connectivity} for finding Strongly Connected Components (SCCs).

\subsubsection{The S-V Algorithm}
\label{sec:eval-scc}

According to the previous discussion, the request-respond channel and the scatter-combine channel are applicable in the algorithm implementation.
We thus have four S-V programs in our system covering all the combination of the two optimized channels.
For comparison, we also give the result of our best-effort implementation in \PP{}'s \textit{basic} and \textit{reqresp} mode.

The results are presented in \autoref{tab:sv-results}.
As expected, the basic version (program 2) without using any specialized channel is the slowest, and the fully optimized version (program 5) takes only half of the execution time.
Furthermore, using either of the request-respond channel (program 3) or the scatter-combine channel (program 4) can lead to a decent improvement, but which one is more effective actually depends on the input graph.
Then, even with the request-respond paradigm, \PP{} is slower than our unoptimized version on Twitter, and is only as good as it on Facebook.

\textbf{Analysis.}
According to the average degree in \autoref{tab:datasets}, Twitter is much denser than Facebook, so the neighborhood communication actually dominates the algorithm.
In such case, the scatter-combine version (program 4) works significantly better than the request-respond version (program 3).
We can clearly see that even using the request-respond optimization, the algorithm is still inefficient when the input graph is dense, since the redundancies in the neighborhood communication becomes the major problem.
Instead, combining two different optimizations can make the algorithm work consistently well, regardless of the density of input graph.

\subsubsection{Min-Label Algorithm}
\label{sec:eval-scc}

\begin{table}[t]
\caption{Experiment results of the Min-Label algorithm}
\label{tab:scc-results}
\begin{tabular}{l|c|c|c|c}
\hline
\multicolumn{1}{c}{\multirow{2}{*}{Program}} & \multicolumn{2}{|c}{Wikipedia} & \multicolumn{2}{|c}{Wikipedia (P)} \\
\cline{2-5}
& \small runtime & \small message & \small runtime & \small message \\
\hline\hline
1-pregel+(basic) & 52.15 & 9.85  & 50.51 & 2.70  \\
\hline
2-channel (basic) & 61.89 & 4.98  & 67.84 & 1.29  \\
\hline
3-channel (prop.) & 31.37 & 4.42  & 13.96 & 1.12  \\
\hline
\end{tabular}
\vspace{-2.5ex}
\end{table}

Strongly connected component (SCC) is a fundamental problem in graph theory and it is widely used in practice to reveal the properties of the graphs.
A typical Min-Label algorithm~\cite{connectivity} for finding SCCs in Pregel is already complex which is an iterative algorithm where the main iteration contains four subroutines, including the removal of trivial SCCs, forward/backward label propagation, SCC recognization and relabeling.
The algorithm suffers the problem of low convergence speed.

Our system offers a quick fix to this problem by choosing a \texttt{Propagation} channel for the forward/backward label propagation, which results in a 2x speedup on Wikipedia, and a nearly 4x faster on partitioned Wikipedia (see \autoref{tab:scc-results}).
This optimization is not possible in any of the existing system.

\section{Related Work}
\label{sec:related}

Google's Pregel~\cite{pregel} is the first specific in-memory system for distributed graph processing.
It adopts the Bulk-Synchronous Parallel (BSP) model~\cite{bsp} with explicit messages to let users implement graph algorithms in a vertex-centric way.
The core design of Pregel has been widely adopted by many open-source frameworks~\cite{survey, survey2}, and most of them inherit the monolithic message passing interface, meaning that the messages of different purposes are mixed and indistinguishable for the system.
As an attempt for optimizing communication patterns,
\PP{} extends Pregel with additional interfaces (in particular, the \emph{reqresp} and the \emph{ghost} mode), but it is less flexible since the two modes cannot be composed and adding optimizations is inconvenient.

To support intuitive message slicing in Pregel-like systems, Telos~\cite{telos} proposes a layered architecture where interleaving tasks are implemented as separate \emph{Protocols}, each having a user-defined \texttt{compute()} function with a dedicated message buffer.
However, it lacks an essential feature for optimization that users cannot modify the implementation of the message buffer.
Husky~\cite{husky} is a general-purpose distributed framework with the channel interface, and it supports primitives like \textit{pull}, \textit{push} and \textit{migrate} and \textit{asynchronous updates} to combine the strength of graph-parallel and machine learning systems.
We extend this idea for composing optimizations in graph-parallel system and propose our optimized channels for three common performance issues.

There has been much research studying the optimizations on Pregel-like systems, and our optimized channels draw inspiration from this line of research, such as the sender-side message combining (a.k.a.~vertex-replication, mirroring)~\cite{xpregel,gps,distrgraphlab,pregelplus}, the request-respond paradigm~\cite{pregel}, the block-centric model~\cite{thinkgraph,blogel,goffish} and so on.\
In particular, our scatter-combine channel recognizes the static messaging pattern and reduces the computational cost as well as message size by pre-processing, which is novel and turns out to be effective for com\-mu\-ni\-ca\-tion-intensive algorithms like PageRank and S-V.
We also demonstrate how complex algorithms like S-V and SCC can be optimized by such technique, while most existing systems only focus on rather simple algorithms.

Apart from Pregel, there are graph-parallel systems that use high-level models to organize the computation and communication, which brings more opportunities for optimization.
For example, the Gather-Apply-Scatter (GAS) model (used by GraphLab~\cite{graphlab}, PowerGraph~\cite{powergraph} and PowerLyra~\cite{powerlyra}) is a typical one that describes a vertex-program by three functions, and the scatter-combine model (used by Graphine~\cite{graphine}) fuses the scatter and gather operations, resulting a more compact two-phase model.
Our channel mechanism shares the same spirit;
through the channels, we can equip a system with even more abstractions, so that users can choose whatever suitable for their algorithms.

There are also graph systems using a functional interface with high-level primitives to manipulate the entire graph, such as GraphX~\cite{graphx} (a library on top of Apache Spark~\cite{spark}) and its extension HelP~\cite{help}.
However, their primitives are hard to compose.
Furthermore, experiment results~\cite{husky} show that they are less efficient than other systems even on simple algorithms like PageRank.
Sparse-matrix based frameworks (e.g. the CombBLAS~\cite{combinatorial} and PEGASUS\cite{pegasus}) are also popular for handling graphs which provide linear algebra primitives, but the lack of graph semantics makes it hard for deep optimizations.

\section{Conclusion}
\label{sec:conclusion}

In this paper, we propose to use the channel interface as a replacement of Pregel's message passing and aggregator mechanism, which 
not only structures the communication in Pregel algorithms in an intuitive way, but also 
enables different kinds of optimizations to work together for dealing with different performance issues in the same program.
We also demonstrate how three optimized channels are implemented in this manner.
Experiments show that, our channel-based system along with the current three optimized channels can achieve significantly better performance for a wide spectrum of graph algorithms.

We believe that, having such system with various modular optimizations available can help users quickly build efficient graph applications.
In particular, our methodology of ``optimizing a program by choosing suitable channels'' hopes to change the current situation that different performance issues are addressed by separate systems, so that users can get rid of the steep cost of learning different systems' interfaces or strengths, but enjoy using a single all-around system.
To further make this system easy to use for non-expert users, in the future, we are going to study the compilation from a high-level declarative domain-specific language Palgol~\cite{palgol} to our system.

\bibliographystyle{abbrv}
\bibliography{ref}

\begin{thebibliography}{10}

\bibitem{giraph}
{Apache Giraph}.
\newblock http://giraph.apache.org/.

\bibitem{pregelplus}
{Pregel+}.
\newblock http://www.cse.cuhk.edu.hk/pregelplus/.

\bibitem{xpregel}
N.~T. Bao and T.~Suzumura.
\newblock Towards highly scalable pregel-based graph processing platform with
  x10.
\newblock In {\em Proceedings of the 22nd International Conference on World
  Wide Web}, pages 501--508. ACM, 2013.

\bibitem{combinatorial}
A.~Bulu{\c{c}} and J.~R. Gilbert.
\newblock The combinatorial {BLAS}: Design, implementation, and applications.
\newblock {\em The International Journal of High Performance Computing
  Applications}, 25(4):496--509, 2011.

\bibitem{powerlyra}
R.~Chen, J.~Shi, Y.~Chen, and H.~Chen.
\newblock {PowerLyra}: Differentiated graph computation and partitioning on
  skewed graphs.
\newblock In {\em Proceedings of the Tenth European Conference on Computer
  Systems}, page~1. ACM, 2015.

\bibitem{boruvka}
S.~Chung and A.~Condon.
\newblock Parallel implementation of {Bor\r{u}vka}'s minimum spanning tree
  algorithm.
\newblock In {\em IPPS}, pages 302--308. IEEE, 1996.

\bibitem{disjointset}
H.~N. Gabow and R.~E. Tarjan.
\newblock A linear-time algorithm for a special case of disjoint set union.
\newblock {\em \textit{{J.} {Comput.} System {Sci.}}}, 30(2):209--221, 1985.

\bibitem{powergraph}
J.~E. Gonzalez, Y.~Low, H.~Gu, D.~Bickson, and C.~Guestrin.
\newblock {PowerGraph}: distributed graph-parallel computation on natural
  graphs.
\newblock In {\em OSDI}, pages 17--30, 2012.

\bibitem{pegasus}
U.~Kang, C.~E. Tsourakakis, and C.~Faloutsos.
\newblock Pegasus: A peta-scale graph mining system implementation and
  observations.
\newblock In {\em Proceedings of the Ninth IEEE International Conference on
  Data Mining}, pages 229--238. IEEE, 2009.

\bibitem{metis}
G.~Karypis and V.~Kumar.
\newblock A fast and high quality multilevel scheme for partitioning irregular
  graphs.
\newblock {\em SIAM Journal on scientific Computing}, 20(1):359--392, 1998.

\bibitem{mizan}
Z.~Khayyat, K.~Awara, A.~Alonazi, H.~Jamjoom, D.~Williams, and P.~Kalnis.
\newblock Mizan: a system for dynamic load balancing in large-scale graph
  processing.
\newblock In {\em EuroSys}, pages 169--182. ACM, 2013.

\bibitem{rmat}
F.~Khorasani, R.~Gupta, and L.~N. Bhuyan.
\newblock Scalable simd-efficient graph processing on gpus.
\newblock In {\em Proceedings of the 24th International Conference on Parallel
  Architectures and Compilation Techniques}, PACT '15, pages 39--50, 2015.

\bibitem{distrgraphlab}
Y.~Low, D.~Bickson, J.~Gonzalez, C.~Guestrin, A.~Kyrola, and J.~M. Hellerstein.
\newblock Distributed graphlab: a framework for machine learning and data
  mining in the cloud.
\newblock {\em Proceedings of the VLDB Endowment}, 5(8):716--727, 2012.

\bibitem{graphlab}
Y.~Low, D.~Bickson, J.~Gonzalez, C.~Guestrin, A.~Kyrola, and J.~M. Hellerstein.
\newblock {Distributed GraphLab}: a framework for machine learning and data
  mining in the cloud.
\newblock {\em Proceedings of the VLDB Endowment}, 5(8):716--727, 2012.

\bibitem{telos}
A.~Lulli, P.~Dazzi, L.~Ricci, and E.~Carlini.
\newblock A multi-layer framework for graph processing via overlay composition.
\newblock In {\em European Conference on Parallel Processing}, pages 515--527.
  Springer, 2015.

\bibitem{pregel}
G.~Malewicz, M.~H. Austern, A.~J. Bik, J.~C. Dehnert, I.~Horn, N.~Leiser, and
  G.~Czajkowski.
\newblock Pregel: a system for large-scale graph processing.
\newblock In {\em SIGMOD}, pages 135--146. ACM, 2010.

\bibitem{survey}
R.~R. McCune, T.~Weninger, and G.~Madey.
\newblock Thinking like a vertex: a survey of vertex-centric frameworks for
  large-scale distributed graph processing.
\newblock {\em ACM Computing Surveys (CSUR)}, 48(2):25, 2015.

\bibitem{QuWH12}
L.~Quick, P.~Wilkinson, and D.~Hardcastle.
\newblock Using {Pregel}-like large scale graph processing frameworks for
  social network analysis.
\newblock In {\em ASONAM}, pages 457--463. IEEE, 2012.

\bibitem{gps}
S.~Salihoglu and J.~Widom.
\newblock {GPS}: a graph processing system.
\newblock In {\em SSDBM}, page~22. ACM, 2013.

\bibitem{help}
S.~Salihoglu and J.~Widom.
\newblock {HelP}: High-level primitives for large-scale graph processing.
\newblock In {\em Proceedings of Workshop on GRAph Data management Experiences
  and Systems}, pages 1--6. ACM, 2014.

\bibitem{optimizing}
S.~Salihoglu and J.~Widom.
\newblock Optimizing graph algorithms on {Pregel}-like systems.
\newblock {\em Proceedings of the VLDB Endowment}, 7(7):577--588, 2014.

\bibitem{ShVi82}
Y.~Shiloach and U.~Vishkin.
\newblock An {$O(\log n)$} parallel connectivity algorithm.
\newblock {\em Journal of Algorithms}, 3:57--67, 1982.

\bibitem{goffish}
Y.~Simmhan, A.~Kumbhare, C.~Wickramaarachchi, S.~Nagarkar, S.~Ravi,
  C.~Raghavendra, and V.~Prasanna.
\newblock {Goffish}: A sub-graph centric framework for large-scale graph
  analytics.
\newblock In {\em European Conference on Parallel Processing}, pages 451--462.
  Springer, 2014.

\bibitem{thinkgraph}
Y.~Tian, A.~Balmin, S.~A. Corsten, S.~Tatikonda, and J.~McPherson.
\newblock From think like a vertex to think like a graph.
\newblock {\em Proceedings of the VLDB Endowment}, 7(3):193--204, 2013.

\bibitem{bsp}
L.~G. Valiant.
\newblock A bridging model for parallel computation.
\newblock {\em \textit{{Commun.} ACM}}, 33(8):103--111, 1990.

\bibitem{graphx}
R.~S. Xin, J.~E. Gonzalez, M.~J. Franklin, and I.~Stoica.
\newblock {GraphX}: A resilient distributed graph system on {Spark}.
\newblock In {\em First International Workshop on Graph Data Management
  Experiences and Systems}, page~2. ACM, 2013.

\bibitem{survey2}
D.~Yan, Y.~Bu, Y.~Tian, A.~Deshpande, et~al.
\newblock Big graph analytics platforms.
\newblock {\em Foundations and Trends{\textregistered} in Databases},
  7(1-2):1--195, 2017.

\bibitem{blogel}
D.~Yan, J.~Cheng, Y.~Lu, and W.~Ng.
\newblock Blogel: A block-centric framework for distributed computation on
  real-world graphs.
\newblock {\em Proceedings of the VLDB Endowment}, 7(14):1981--1992, 2014.

\bibitem{effective}
D.~Yan, J.~Cheng, Y.~Lu, and W.~Ng.
\newblock Effective techniques for message reduction and load balancing in
  distributed graph computation.
\newblock In {\em International World Wide Web Conference}, pages 1307--1317.
  ACM, 2015.

\bibitem{connectivity}
D.~Yan, J.~Cheng, K.~Xing, Y.~Lu, W.~Ng, and Y.~Bu.
\newblock Pregel algorithms for graph connectivity problems with performance
  guarantees.
\newblock {\em Proceedings of the VLDB Endowment}, 7(14):1821--1832, 2014.

\bibitem{graphine}
J.~Yan, G.~Tan, Z.~Mo, and N.~Sun.
\newblock Graphine: programming graph-parallel computation of large natural
  graphs for multicore clusters.
\newblock {\em IEEE Transactions on Parallel and Distributed Systems},
  27(6):1647--1659, 2016.

\bibitem{husky}
F.~Yang, J.~Li, and J.~Cheng.
\newblock Husky: Towards a more efficient and expressive distributed computing
  framework.
\newblock {\em Proceedings of the VLDB Endowment}, 9(5):420--431, 2016.

\bibitem{spark}
M.~Zaharia, M.~Chowdhury, M.~J. Franklin, S.~Shenker, and I.~Stoica.
\newblock Spark: Cluster computing with working sets.
\newblock {\em HotCloud}, 10(10-10):95, 2010.

\bibitem{palgol}
Y.~Zhang, H.-S. Ko, and Z.~Hu.
\newblock Palgol: A high-level {DSL} for vertex-centric graph processing with
  remote data access.
\newblock In {\em Proceedings of the 15th Asian Symposium on Programming
  Languages and Systems}, pages 301--320. Springer, 2017.

\end{thebibliography}

\end{document}